\documentclass[%
 reprint,
 amsmath,amssymb,
 aps,
]{revtex4-2}

\usepackage{graphicx}
\usepackage{dcolumn}
\usepackage{bm}

\begin{document}

\preprint{APS/123-QED}

\title{Manifestation of strong and ultra-strong coupling in high-order correlation function}

\author{A. S. Belashov}
\author{E. S. Andrianov}
\author{A. A. Zyablovsky}
\email{zyablovskiy@mail.ru}

\affiliation{Moscow Institute of Physics and Technology, 9 Institutskiy pereulok, Moscow, Russia, 141700}
\affiliation{Dukhov Research Institute of Automatics, 22 Sushchevskaya, Moscow, Russia, 127055}

\date{\today}

\begin{abstract}
  Strong and ultra-strong coupling in “cavity - single atom” system are of great interest for both fundamental and applied physics. It is considered that the increase in the coupling strength between a cavity mode and an atom leads, first, to transition from weak to strong coupling and, second, to ultra-strong coupling regime. In this letter, we refute this common opinion and demonstrate that the transitions between the coupling regimes occur in different sequence for the correlations' functions of different orders. We show that for n-th order correlations' functions, the transition to the strong coupling regime requires the coupling strength approximately by $n^{2/3}$ times greater than the one for first order correlations' functions. In contrast, the transition to the ultra-strong coupling regime manifests in the dynamics of n-th order correlations' functions at the less coupling strength than in the dynamics of first order correlations' functions. As a result, there is the order of correlations' functions, above which the increase in the coupling strength leads, first, to the transition from the weak coupling first to the ultra-strong coupling regime, and second to the strong coupling regime. We argue that the measurement of high orders correlations' functions makes it possible to observe the ultra-strong coupling in “cavity mode – single atom” when the coupling strength is much less than one tenth of the oscillation frequency.
\end{abstract}

\maketitle

\textit{Introduction.} Creation of strong and ultra-strong coupling in “cavity - single atom” systems is a challenge in the quantum optics \cite{RN3, RN7, RN1, RN5, RN2, RN4, RN6}. In the conventional condition, the coupling with a single atom is weak and the system is said to be in a weak coupling regime \cite{RN3, RN2}. The transition to a strong coupling regime occurs when the coupling strength exceeds relaxation rates in the system \cite{RN3, RN2}. In this regime, the Rabi splitting manifests itself as a change in the spectral and coherent properties of radiation \cite{RN3, RN8, RN13, RN12, RN10, RN11, RN14, RN2, RN9}. The strong coupling systems are used to create the quantum light sources \cite{RN16, RN15, RN17} and lasers without population inversion \cite{RN18, RN19, RN20}, to control the chemical reaction rates \cite{RN21, RN22, RN23, RN24, RN26, RN25} and to obtain polariton Bose-Einstein condensates \cite{RN27, RN28, RN30, RN29}, etc. To achieve the strong coupling, both a decrease in the relaxation rates and an increase in the coupling strength can be used \cite{RN3}. Both high-Q cavities and the structure cooling \cite{RN32, RN33, RN31, RN6} are used to diminish the relaxation rates and to achieve the strong coupling in “cavity - single atom” systems. In turn, the coupling strength of a cavity mode with a single atom determines the dipole moment of the atom and the electric field amplitude per one quantum at the atom’s position \cite{RN34}. To increase the coupling strength via the enhancement of the electric field, the structure with subwavelength localization of the light are designed \cite{RN3, RN33, RN2, RN9}. Currently, the developed approaches make it possible to achieve the strong coupling in the in “cavity - single atom” systems \cite{RN3, RN7, RN5, RN4, RN6}, but it remains to be a difficult problem.

The transition to the ultra-strong coupling occurs when the coupling strength exceeds one tenth of the system frequency \cite{RN35, RN1}. The observation of such a transition requires an even greater increase in the coupling strength and, as a result, the use of even more complex technological approaches. In the ultra-strong coupling regime, the counter-rotating wave and diamagnetic terms begin to play an important role \cite{RN35, RN1}. These terms modify the energy states of the system \cite{RN40, RN35, RN38, RN1, RN36, RN37, RN39}, in particular, leading to the fact that the state with zero number of excitation ceases to be the ground state of the system \cite{RN1, RN41}. The change in the energy states endows the ultra-strong coupled systems with unique properties \cite{RN35, RN1} and qualitatively modify the system dynamics.

In this letter, we consider a “cavity mode – single atom” system. Using the master equation for the density matrix, we derive the equations for correlation functions of different orders. We show that the relaxation rate of n-th order correlation function, $\left\langle {{\hat a}^{\dag n}{{\hat a}^n}} \right\rangle$, is $n$ times greater than the one of the first order correlation function, $\left\langle {{{\hat a}^\dag }\hat a} \right\rangle$. While, the frequency of Rabi oscillations for n-th order correlations functions is ${n^{1/3}}$ times greater the coupling strength than the one for $\left\langle {{{\hat a}^\dag }\hat a} \right\rangle$. Thus, the relaxation rates increase with ${n^{2/3}}$ faster than the frequency of Rabi oscillations. Therefore, for n-th order correlation function the transition to the strong coupling regime requires the coupling strength approximately by ${n^{2/3}}$ times greater than one necessary for the transition for the first order correlation function. At the same time, the increase in the oscillations frequency simplifies the transition to the ultra-strong coupling regime. For high order correlations functions, the transition to the ultra-strong coupling regime occurs at less coupling strengths than the one for the first order correlation function. As a result, there is a number n such that for n-th and higher order correlation functions, the system first passes from the weak coupling regime to the ultra-strong coupling regime, and only then to the strong coupling regime. We confirm this conclusion by direct numerical simulation of the equations for the correlation functions of different orders.

Thus, we conclude that the coupling strengths at which the transitions between the weak, strong and ultra-strong-coupling regimes occur depend on the order of correlation functions. The measurement of high order correlation functions makes it possible to observe the transition to the ultra-strong coupling at less coupling strengths than one tenth of the cavity frequency. This result could simplify observations of ultra-strong coupling in “cavity mode – single atom” systems.

\textit{System under consideration.} We consider a system consisting of a single two-level atom and optical mode in a cavity, which resonantly interact with each other. We use the following Hamiltonian for description of this system (for simplicity, $\hbar  = 1$) \cite{RN1}:

\begin{equation}
\hat H = {\omega _0}{\hat a^\dag }\hat a + {\omega _0}{\hat \sigma ^\dag }\hat \sigma  + \Omega \left( {{{\hat a}^\dag } + \hat a} \right)\left( {{{\hat \sigma }^\dag } + \hat \sigma } \right) + {D_a}{\left( {{{\hat a}^\dag } + \hat a} \right)^2}
\label{eq:refname1}
\end{equation}

Here $\hat a$ and ${\hat a^\dag }$ are the annihilation and creation operators of the photons in the optical mode, $\left[ {\hat a,{{\hat a}^\dag }} \right] = 1$. $\hat \sigma$ and ${\hat \sigma ^\dag }$ are the lowering and raising operators for the two-level atom, $\left\{ {\hat \sigma ,{{\hat \sigma }^\dag }} \right\} = 1$. The frequency of the optical mode, ${\omega _0}$, is equal to the frequency of transition in the two-level atom. $\Omega $ is a coupling strength between the optical mode and the atom. The term $\Omega \left( {{{\hat a}^\dag } + \hat a} \right)\left( {{{\hat \sigma }^\dag } + \hat \sigma } \right)$ in the Hamiltonian~(\ref{eq:refname1}) can be rewritten as a sum of two terms: $\Omega \left( {{{\hat a}^\dag }\hat \sigma  + \hat a{{\hat \sigma }^\dag }} \right)$ and $\Omega \left( {{{\hat a}^\dag }{{\hat \sigma }^\dag } + \hat a\hat \sigma } \right)$. The term $\Omega \left( {{{\hat a}^\dag }\hat \sigma  + \hat a{{\hat \sigma }^\dag }} \right)$ is used to describe the interaction between the optical mode and atom in the rotating wave approximation \cite{RN1}. It is considered that this approximation is valid when $\Omega  <  < {\omega _0}$. At the same time, $\Omega \left( {{{\hat a}^\dag }{{\hat \sigma }^\dag } + \hat a\hat \sigma } \right)$ and ${D_a}{\left( {{{\hat a}^\dag } + \hat a} \right)^2}$ are the counter-rotating wave and diamagnetic terms, which must be taken into account when $\Omega  \geqslant 0.1\,{\omega _0}$ \cite{RN1}. To guarantee the set of eigenenergies is bound from below, it is necessary that ${D_a} \geqslant {\Omega ^2}/2{\omega _0}$ \cite{RN1}, which we use in the following consideration.

We use the master equation for the density matrix in the Lindblad form \cite{RN42, RN43} to describe the relaxation processes in the system under consideration:

\begin{equation}
\hat \dot \rho  =  - i\left[ {\hat H,\hat \rho } \right] + {\hat L_a}\left[ {\hat \rho } \right] + {\hat L_D}\left[ {\hat \rho } \right] + {\hat L_P}\left[ {\hat \rho } \right] + {\hat L_\sigma }\left[ {\hat \rho } \right]
\label{eq:refname2}
\end{equation}

Here ${\hat L_a}\left[ {\hat \rho } \right] = \frac{{{\gamma _a}}}{2}\left( {2\hat a\hat \rho {{\hat a}^\dag } - {{\hat a}^\dag }\hat a\hat \rho  - \hat \rho {{\hat a}^\dag }\hat a} \right)$ is the Lindblad superoperator describing the relaxation of the electromagnetic field in the optical mode; ${\hat L_D}\left[ {\hat \rho } \right] = \frac{{{\gamma _D}}}{2}\left( {2\hat \sigma \hat \rho {{\hat \sigma }^\dag } - {{\hat \sigma }^\dag }\hat \sigma \hat \rho  - \hat \rho {{\hat \sigma }^\dag }\hat \sigma } \right)$ describes the relaxation of the population inversion in the two-level atom; ${\hat L_P}\left[ {\hat \rho } \right] = \frac{{{\gamma _P}}}{2}\left( {2{{\hat \sigma }^\dag }\hat \rho \hat \sigma  - \hat \sigma {{\hat \sigma }^\dag }\hat \rho  - \hat \rho \hat \sigma {{\hat \sigma }^\dag }} \right)$ describes the pumping of the two-level atom; ${\hat L_\sigma }\left[ {\hat \rho } \right] = {\gamma _\sigma }\left( {\hat D\hat \rho \hat D - \hat \rho } \right)$ describes the dephasing processes in the two-level atom, where $\hat D = {\hat \sigma ^\dag }\hat \sigma  - \hat \sigma {\hat \sigma ^\dag }$  is an operator of the population inversion in the two-level atom.

\textit{The equations for average values of operators.} Using the master equation for the density matrix~(\ref{eq:refname2}), we derive the equations for the average values of the operators ${\left( {{{\hat a}^\dag }} \right)^n}{\hat a^m}\hat d$, where $\hat d$ denotes the one of the four operators: $\hat 1$, $\hat \sigma $, ${\hat \sigma ^\dag }$ and ${\hat \sigma ^\dag }\hat \sigma $. To start, we consider the equations without taking into account the contribution of the counter-rotating wave and diamagnetic terms in the Hamiltonian~(\ref{eq:refname1}). In this case, the differential equations for $\left\langle {{{\left( {{{\hat a}^\dag }} \right)}^n}{{\hat a}^m}\hat d} \right\rangle $ contain the averages with greater number of the excitation, such as $\left\langle {{{\left( {{{\hat a}^\dag }} \right)}^{n + 1}}{{\hat a}^m}\hat d} \right\rangle$, $\left\langle {{{\left( {{{\hat a}^\dag }} \right)}^n}{{\hat a}^{m + 1}}\hat d} \right\rangle$, $\left\langle {{{\left( {{{\hat a}^\dag }} \right)}^{n + 1}}{{\hat a}^{m + 1}}\hat d} \right\rangle$, which, in turn, contain the averages with more greater number of the excitation. As a result, the system of equations on operator averages is infinite. To obtain the closed system of equations, we consider that the number of excitation in the optical mode is limited by a certain value $N$. We choose $N$ so large that its further increase does not affect the dynamics of considered correlation functions. In this approximation, all averages of operators containing powers $\hat a$ and ${\hat a^\dag }$ greater than $N$ are zero. As a result, we obtain the closed system of equations. The system of equations is divided into subsystems, each of which contains only averages of the operators with a certain difference, $J = {N_ + } - {N_ - }$, of numbers of the atomic raising and field creation operators, ${N_ + }$, and the atomic lowering and field annihilation operators, ${N_ - }$. In particular, the subsystem with $J = 0$ contains the averages of operators as $\left\langle {{{\hat a}^\dag }\hat a} \right\rangle$, $\left\langle {{{\hat \sigma }^\dag }\hat \sigma } \right\rangle$, etc. The corresponding equations for the average of the operators with $J=0$ have form

\begin{equation}
\begin{gathered}
\frac{d}{{dt}}\left\langle {{\hat a}^{\dag n}{{\hat a}^n}} \right\rangle  =  - 2{\gamma _a}n\left\langle {{\hat a}^{\dag n}{{\hat a}^n}} \right\rangle \\
+ i\,\Omega \,n\left( {\left\langle {{\hat a}^{\dag n}{{\hat a}^{n - 1}}\hat \sigma } \right\rangle  - \left\langle {{{\hat a}^{\dag (n - 1)}}{{\hat a}^n}{{\hat \sigma }^\dag }} \right\rangle } \right)
\end{gathered}
\label{eq:refname3}
\end{equation}

\begin{equation}
\begin{gathered}
  \frac{d}{{dt}}\left\langle {{\hat a}^{\dag (n - 1)}{{\hat a}^{n - 1}}{{\hat \sigma }^\dag }\hat \sigma } \right\rangle  = 2{\gamma _P}\left\langle {{\hat a}^{\dag (n - 1)}{{\hat a}^{n - 1}}} \right\rangle \\
  - \left( {2\left( {n - 1} \right){\gamma _a} + 2{\gamma _P} + 2{\gamma _D}} \right)\left\langle {{\hat a}^{\dag (n - 1)}{{\hat a}^{n - 1}}{{\hat \sigma }^\dag }\hat \sigma } \right\rangle  -  \hfill \\
  i\,\Omega \left( {\left\langle {{\hat a}^{\dag n}{{\hat a}^{n - 1}}\hat \sigma } \right\rangle  - \left\langle {{\hat a}^{\dag (n - 1)}{{\hat a}^n}{{\hat \sigma }^\dag }} \right\rangle } \right) \hfill \\ 
\end{gathered}
\label{eq:refname4}
\end{equation}

\begin{equation}
\begin{gathered}
  \frac{d}{{dt}}\left\langle {{\hat a}^{\dag n}{{\hat a}^{n - 1}}\hat \sigma } \right\rangle  = \\
  - \left( {\left( {2 n - 1} \right){\gamma _a} + {\gamma _\sigma } + {\gamma _P} + {\gamma _D}} \right)\left\langle {{\hat a}^{\dag n}{{\hat a}^{n - 1}}\hat \sigma } \right\rangle + i\,\Omega \left\langle {{\hat a}^{\dag n}{{\hat a}^n}} \right\rangle  \hfill \\
  - 2i\,\Omega \left\langle {{\hat a}^{\dag n}{{\hat a}^n}{{\hat \sigma }^\dag }\hat \sigma } \right\rangle  - i\,\Omega \,n\left\langle {{\hat a}^{\dag (n - 1)}{{\hat a}^{n - 1}}{{\hat \sigma }^\dag }\hat \sigma } \right\rangle  \hfill \\ 
\end{gathered}
\label{eq:refname5}
\end{equation}

The quantity $\left\langle {{\hat a}^{\dag (n - 1)}{{\hat a}^n}{{\hat \sigma }^\dag }} \right\rangle$ is calculated as a complex conjugate of $\left\langle {{\hat a}^{\dag n}{{\hat a}^{n - 1}}\hat \sigma } \right\rangle$.

\textit{Eigenfrequencies of the system.} The Eqns.~(\ref{eq:refname3})–(\ref{eq:refname5}) for correlation functions of n-th order are not closed because they include the correlation functions of the correlation functions of (n-1)-th and (n+1)-th orders, $\left\langle {{\hat a}^{\dag (n - 1)}{{\hat a}^{n - 1}}} \right\rangle$ and $\left\langle {{\hat a}^{\dag n}{{\hat a}^n}{{\hat \sigma }^\dag }\hat \sigma } \right\rangle$, respectively. For this reason, the calculation of the dynamics of n-th order correlation functions requires simulation of the entire chain of equations. The characteristic oscillation frequencies and relaxation rates of the n-th order correlation functions can be estimated from the Eqns.~(\ref{eq:refname3})–(\ref{eq:refname5}). It is seen that the relaxation rates $2{\gamma _a}n$ in the Eqns.~(\ref{eq:refname3})–(\ref{eq:refname5}) increase linearly with $n$. At the same time, the coupling strengths between the correlations' functions of n-th order depend on $n$. As a result, the frequencies of Rabi oscillations are also depend on $n$. To estimate the oscillation frequencies, we consider the Eqns.~(\ref{eq:refname3})–(\ref{eq:refname5}) leaving only the leading terms with respect to $n$ (in this consideration the equations for correlations' functions of n-th order do not contain the correlations' functions of other orders) and then calculate the eigenfrequencies (real part of eigenvalues) and relaxation rates (imaginary parts of eigenvalues). Our calculations show that the relaxation rates approximately linearly increase with $n$ ($\sim 2{\gamma _a}n$). At the same time, the eigenfrequencies increase proportionally to ${n^{1/3}}$ ($\sim {n^{1/3}}\,\Omega$).

The ratio of the characteristic oscillation frequencies and the relaxation rates determines the regime (weak, strong, or ultra-strong) taking place in the system. Since, the relaxation rates and the characteristic oscillation frequencies have different dependences on $n$, we can conclude that the transition from weak to strong coupling and then to ultra-strong coupling occurs at the different values of ${\gamma _a}$ and $\Omega$ for the correlation functions of different orders.

\textit{Transition from weak to strong coupling regime.} The weak coupling regime takes place when the coupling strength is less than the relaxation rates. In this regime, the exponential decay in the system occurs. The transition from weak to strong coupling regime is characterized by an appearance of the Rabi oscillations in the system dynamics. The condition for the observation of the Rabi oscillations is that the oscillation period is less than the decay time. Therefore, the strong coupling regime takes place when the frequency of the Rabi oscillations exceeds all relaxation rates.

In the previous section, it follows that the oscillation frequency is proportional to ${n^{1/3}}$, where $n$ is the order of the correlation function (${\Omega _n} \sim {n^{1/3}}\,\Omega $). At the same time, the relaxation rates increase linearly with $n$ (${\gamma _n} \sim 2{\gamma _a}n$). Using the condition for the strong coupling regime

\begin{equation}
{\Omega _n} > {\gamma _n}\,\,\,\,\, \Leftrightarrow {n^{1/3}}\,\Omega  > 2{\gamma _a}n
\label{eq:refname6}
\end{equation}

we obtain that the value of the coupling strength ${\Omega _{SC}}$, at which the transition to the strong coupling regime occurs, depends on $n$ as:

\begin{equation}
{\Omega _{SC}} = 2{\gamma _a}\,{n^{2/3}}
\label{eq:refname7}
\end{equation}

\begin{figure}[htbp]
\centering\includegraphics[width=\linewidth]{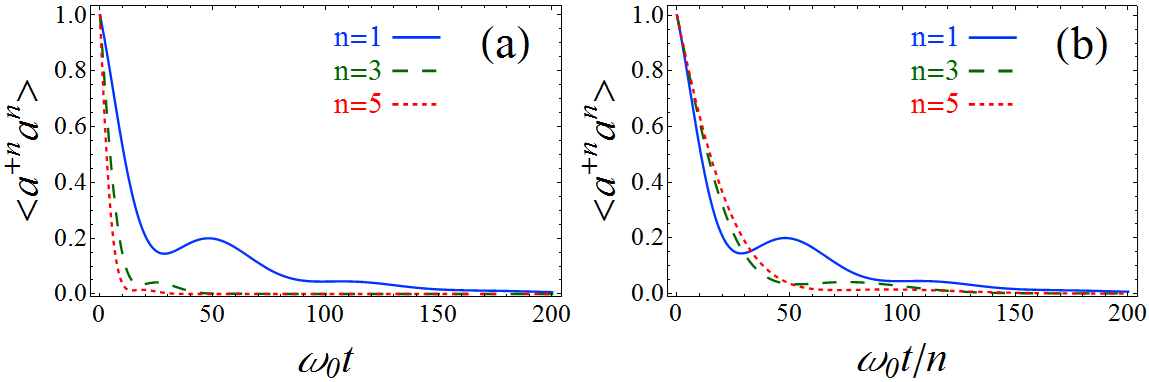}
\caption{The temporal dynamic of the first order correlation function, $\left\langle {{{\hat a}^\dag }\hat a} \right\rangle$, (solid blue line); the third order correlation function, $\left\langle {{{\hat a}^{\dag 3}}{{\hat a}^3}} \right\rangle$, (dashed green line); the fifth order correlation function, $\left\langle {{\hat a}^{\dag 5}{{\hat a}^5}} \right\rangle$, (dotted red line). (a) The dependences for all the correlations' functions are plotted on the same time scale; (b) the dependences for the correlations' functions are plotted on time scales normalized to $n$. ${\gamma _P}/{\gamma _D} = 0$; $\Omega  = 0.05\,{\omega _0}$.}
\label{fig1}
\end{figure}

Therefore, to observe the strong coupling regime in the behavior of the high orders' correlation functions, it is necessary for achieving the greater coupling strength between the optical mode and the atom. To prove this statement, we calculate from numerical simulation of Eqns.~(\ref{eq:refname3})–(\ref{eq:refname5}) the temporal dynamics of the correlation functions of the different orders [Figure~\ref{fig1}]. It is seen that the Rabi oscillations are more clearly visible in the dynamics of the first order correlation function. When the order of the correlation function increases, the Rabi oscillations become less pronounced. It confirms that for the high orders' correlation functions the transition to strong coupling occurs at greater coupling strength between the mode and atom.

\textit{Transition from weak to strong coupling regime.} 
Growth of the coupling strength increases the influence of the counter-rotating wave and diamagnetic terms on the system dynamics. These terms change the energy levels of the system \cite{RN1} that manifests in the system spectrum. The influence of the counter-rotating wave and diamagnetic terms can be estimated by the Bloch-Siegert shift \cite{RN44}, which determines the change in the oscillation frequency and is proportional to ${\Omega ^2}/{\omega _0}$. The frequency shift in the rotating wave approximation is proportional to $\Omega $. Thereby, it is the ratio of $\Omega /{\omega _0}$ that determines the influence of the counter-rotating wave and diamagnetic terms. It is considered \cite{RN1} that when $\Omega /{\omega _0} > 0.1$, the system is in the ultra-strong coupling regime. In this regime, the counter-rotating wave and diamagnetic terms have a noticeable effect on the system dynamics \cite{RN1}.
Taking into account that the oscillation frequency increases with $n$ and the condition for the ultra-strong coupling regime \cite{RN1} is 

\begin{equation}
{\Omega _n} > 0.1\,{\omega _0}
\label{eq:refname8}
\end{equation}
we can conjecture that the value of the coupling strength, at which the transition to the ultra-strong coupling regime occurs, ${\Omega _{USC}}$, decreases with $n$ (the dependence of $\Omega_n$ on $n$ for the systems with the counter-rotating wave and diamagnetic terms can differ from the one estimated by the Eqns.~(\ref{eq:refname3})–(\ref{eq:refname5})). We can surmise that the ultra-strong coupling regime is observed in the correlations' functions of high orders at lower coupling strength than the one in the correlation functions of low orders. To prove this statement, we derive the equations for averages the operators taking into account the counter-rotating wave and diamagnetic terms in the Hamiltonian (\ref{eq:refname1}). These equations contain an explicit dependency on $\omega_0$.

\begin{figure}[htbp]
\centering\includegraphics[width=\linewidth]{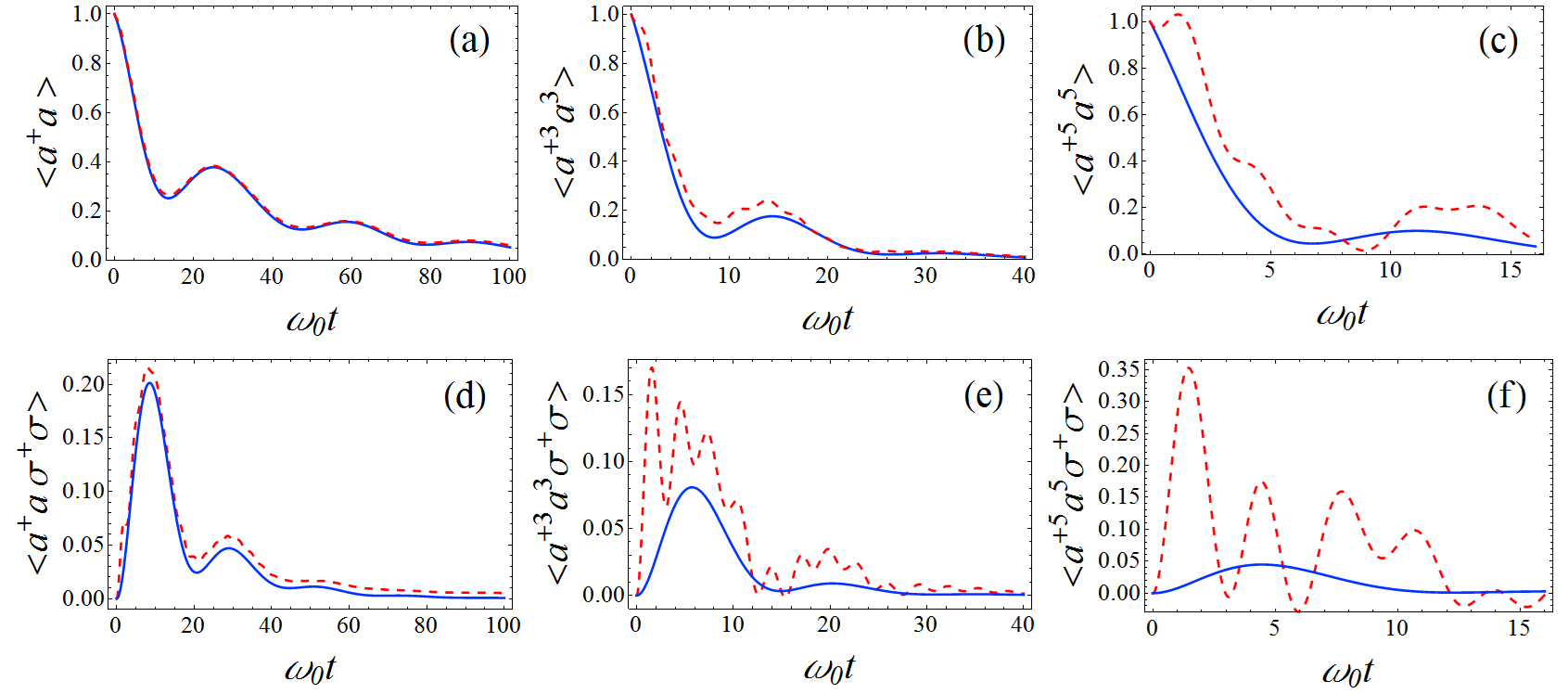}
\caption{The temporal dependences of the correlation functions:   $\left\langle {{{\hat a}^\dag }\hat a} \right\rangle$ (a); $\left\langle {{\hat a}^{\dag 3}{{\hat a}^3}} \right\rangle$ (b); $\left\langle {{\hat a}^{\dag 5}{{\hat a}^5}} \right\rangle $ (c); $\left\langle {{{\hat a}^\dag }\hat a{{\hat \sigma }^\dag }\hat \sigma } \right\rangle$ (d); $\left\langle {{\hat a}^{\dag 3}{{\hat a}^3}{{\hat \sigma }^\dag }\hat \sigma } \right\rangle $ (e); $\left\langle {{\hat a}^{\dag 5}{{\hat a}^5}{{\hat \sigma }^\dag }\hat \sigma } \right\rangle$ (f). The blue lines are calculated without taking into account the counter-rotating wave and diamagnetic terms. The red lines are calculated with taking into account the counter-rotating wave and diamagnetic terms.}
\label{fig2}
\end{figure}

We simulate the equations taking into account the counter-rotating wave and diamagnetic terms and the Eqns.~(\ref{eq:refname3})–(\ref{eq:refname5}) that do not take into account such terms. We calculate the temporal dynamics of the correlation functions of different orders [Figure~\ref{fig2}]. It is seen that for the correlations' functions of high orders the differences in the system dynamics calculated with and without taking into account the counter-rotating wave and diamagnetic terms are clearly visible [Figure~\ref{fig2}b, c, e, f]. While, for the correlations' functions of low orders these differences are invisible [Figure~\ref{fig2}a, d]. It confirms that for the correlations' functions of high orders the transition to ultra-strong coupling occurs at lower coupling strength than the one for the correlations' functions of low orders.

To characterize numerically the difference between the system dynamics calculated with and without taking into account the counter-rotating wave and diamagnetic terms, we introduce the following quantity

\begin{equation}
\begin{gathered}
\Delta \left( n \right) = \frac{{\int {{{\left| {{{\left\langle {{\hat a}^{\dag n}{{\hat a}^n}} \right\rangle }_{with}}} \right|}^2}} \,dt - \int {{{\left| {{{\left\langle {{\hat a}^{\dag n}{{\hat a}^n}} \right\rangle }_{without}}} \right|}^2}} \,dt}}{{\int {{{\left| {{{\left\langle {{\hat a}^{\dag n}{{\hat a}^n}} \right\rangle }_{without}}} \right|}^2}} \,dt}}
\end{gathered}
\label{eq:refname9}
\end{equation}
where ${\left\langle {{\hat a}^{\dag n}{{\hat a}^n}} \right\rangle _{with}}$ and ${\left\langle {{\hat a}^{\dag n}{{\hat a}^n}} \right\rangle _{without}}$ are the n-th order correlation functions calculated with and without taking into account the counter-rotating wave and diamagnetic terms. The value $\Delta \left( n \right) \gg 1$ means that counter-rotating and diamagnetic terms give the main contribution to the system dynamics.

\begin{figure}[htbp]
\centering\includegraphics[width=\linewidth]{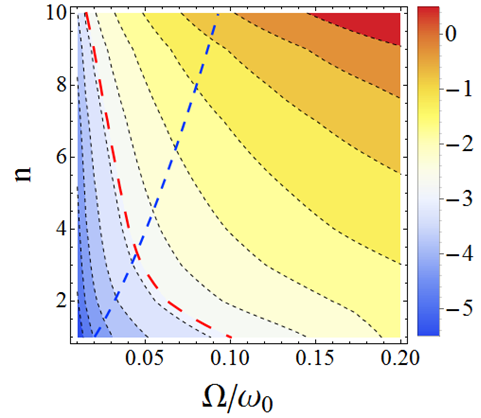}
\caption{The dependence of $\ln \Delta \left( n \right)$ (see Eq.~(\ref{eq:refname9})) on $n$ and $\Omega$. The dashed blue line shows the curve~(\ref{eq:refname7}) (the boundary of the strong coupling regime); the dashed red line shows the boundary of the ultra-strong coupling regime.}
\label{fig3}
\end{figure}

We calculate $\Delta \left( n \right)$ for the different $n$ and $\Omega$ [Figure~\ref{fig3}]. The difference between the system dynamics calculated with and without taking into account the counter-rotating wave and diamagnetic terms grows with the increase of both $n$ and $\Omega$. Thus, the transition to the ultra-strong coupling can occur with the increase of the coupling strength as well as the order of the measured correlation functions. Our numerical calculations show that the coupling strength, at which the transition to the ultra-strong coupling regime occurs, ${\Omega _{USC}}$, decreases with $n$ as [see dashed red line in Figure~\ref{fig3}]. It is seen that, for example, the transition to the ultra-strong coupling regime for tenth order correlation function occurs at $\Omega  \approx 0.015\,{\omega _0} <  < 0.1\,{\omega _0}$. Therefore, we can conclude that the measurement of the correlations' functions of high orders simplifies the observation of the ultra-strong coupling effects.

Note that there is value of $n$ such that for n-th and higher order correlation functions, the system is in the ultra-strong coupling regime, but not in the strong coupling regime [see curves in Figure~\ref{fig3}].

\textit{Transition from weak to strong coupling regime.} In conclusion, we show that the coupling strengths between the mode and atom at which the transitions between the weak, strong and ultra-strong-coupling regimes occur depend on the order of correlation function. For correlations' functions of low orders the increase in the coupling strength leads to the transition from the weak coupling regime to the strong coupling one and then to the ultra-strong coupling regime. While for the correlations' functions of high orders, the same increase in the coupling strength leads to the transition from weak coupling regime to the ultra-strong coupling regime and only then to the strong coupling regime. Thus, the transitions between coupling regimes occur in the different manner for the correlations' functions of different orders. We demonstrate that the measurement of the correlations' functions of high orders makes it possible to observe the transition to the ultra-strong coupling at less coupling strengths than one tenth of the system eigenfrequencies. This can simplify the observations of the ultra-strong coupling in “cavity mode – single atom” systems.

\
\section*{acknowledgement}

The study was financially supported by a Grant from Russian Science Foundation (Project No. 20-72-10057).

\nocite{*}

\bibliography{apssamp}

\end{document}